\documentclass[a4paper,11pt]{article}
\usepackage{harvard}
\usepackage{url}
\usepackage[dvips]{graphics}
\usepackage{epsfig}
\usepackage{psfrag}
\usepackage[psamsfonts]{amssymb}
\usepackage{amsmath}
\usepackage{indentfirst}
\usepackage{amssymb}

\newcommand \be{\begin{equation}}
\newcommand \bea{\begin{eqnarray}}
\newcommand \ee{\end{equation}}
\newcommand \eea{\end{eqnarray}}

\renewcommand{\epsilon}{\varepsilon}

\begin{document}

\title{Theory of Zipf's Law and of General Power Law Distributions with Gibrat's law of Proportional Growth
\thanks{The authors acknowledge helpful discussions and exchanges with Xavier Gabaix.}
} \thispagestyle{empty}
\author{A. Saichev$^{1,2}$, Y. Malevergne$^{3,4}$,  and D. Sornette$^{1,5}$ \thanks{Corresponding author: D. Sornette,  Chair of Entrepreneurial Risks - ETH Z\"urich, KPL F 38.2, Kreuzplatz 5, 8032 Z\"urich ,
Switzerland. Tel: +41 44 63 28917   , Fax: +41 44 632 19 14.}
\\ \\
$^1$ ETH Z\"urich -- Department of Management, Technology and Economics\\ 8032 Z\"urich, Switzerland\\
$^2$ Mathematical Department, Nizhny Novgorod State University, \\ Gagarin Prospekt
23., Nizhny Novgorod, 603950, Russia\\
$^3$ University of Saint-Etienne -- ISEAG, France\\
$^4$ EM-Lyon Business School \\ Department Economics, Finance and Control, France\\
$^5$ Swiss Finance Institute\\ c/o University of Geneva, 40 blvd. Du Pont d®Arve \\CH 1211 Geneva 4, Switzerland\\
saichev@hotmail.com, ymalevergne@ethz.ch and dsornette@ethz.ch }

\date{}
\maketitle

\begin{abstract}

We summarize a book under publication with the above title written by
the three present authors, on the theory
of Zipf's law, and more generally of power laws, driven by the 
mechanism of proportional growth. The preprint is available upon
request from the authors.

For clarity, consistence of language and conciseness, we discuss the origin and
conditions of the validity of Zipf's law using the terminology of firms' asset values. 
We use firms at the
entities whose size distributions are to be explained. It should be noted, however,
that most of the relations discussed in this book, especially the intimate
connection between Zipf's and Gilbrat's laws, underlie Zipf's law in diverse
scientific areas. The same models and variations thereof can be straightforwardly
applied to any of the other domains of application.

\end{abstract}
\vspace{2cm}

 \vspace{0.5cm}



\section*{Executive summary}
 
Zipf's law is one of the few quantitative reproducible regularities found in
economics. It states that, for most countries, the size distributions of city sizes
and of firms (with additional examples found in many other scientific fields) are
power laws with a specific exponent: the number of cities and of firms with size
greater than $S$ is inversely proportional to $S$. 

Most explanations start with
Gibrat's law of proportional growth but need to incorporate additional constraints
and ingredients introducing deviations from it. 

Here, we present a general
theoretical derivation of Zipf's law, providing a synthesis and extension of
previous approaches. First, we show that combining Gibrat's law at all firm levels
with random processes of firms' births and deaths yield Zipf's law under a
``balance'' condition between firm growth and their death rate. 

We find that
Gibrat's law of proportionate growth does not need to be strictly satisfied. As
long as the volatility of firms' sizes increases asymptotically proportionally to
the size of the firm and that the instantaneous growth rate increases not faster
than the volatility, the distribution of firm sizes follows Zipf's law. This
suggests that the occurrence of very large firms in the distribution of firm sizes
described by Zipf's law is more a consequence of random growth than systematic
returns: in particular for large firms, volatility must dominate over the
instantaneous growth rate. 

We develop the theoretical framework to take into
account 
\begin{enumerate}
\item time-varying firm creation, 
\item firms' exit resulting from both a lack
of sufficient capital and sudden external shocks, and
\item  the coupling between firms'
birth rate and the growth of the value of the population of firms. 
\end{enumerate}

We predict
deviations from Zipf's law under a variety of circumstances, for instance when the
balance between the birth rate, the non-stochastic growth rate and the death rate
is not fulfilled, providing a framework for identifying the possible origin(s) of
the many reports of  deviations from the pure Zipf's law. 
The tail index that characterizes the hyperbolic decay of the distribution is found to depend on several characteristics of the economic environment. Amongst others, the average growth rate of firms' asset value, the rate of firms' birth and the hazard rate of a firm's sudden death have a direct impact on the value of the tail index.

Reciprocally, deviations
from Zipf's law in a given economy provides a diagnostic, suggesting possible
policy corrections.  The results obtained here are general and provide an
underpinning for understanding and quantifying Zipf's law and the power law
distribution of sizes found in many fields.

A general result unraveled by our study is that Zipf's law is obtained if and only if a balanced condition is fulfilled: the sum of all the mechanisms responsible for the growth and/or decline of firms must vanish on average. Any departure from this requirement yields a departure of the tail index from its canonical value $m=1$. This result can allow one to understand why different tail indexes are reported in the literature for different countries around the world. However, the reasons that underpin the validity of the balance condition are not yet clear. No economic law can justify why all these mechanisms should almost exactly compensate one another. In the absence of such economic argument, one has to resort to Gabaix's explanation based upon the idea that, in order to make stationary the distribution of firm's sizes, one has to first remove the impact of the overall economy on the growth of each individual firm. Therefore, since the overall economy grows at the same rate as each individual firm, on average, the balance condition is satisfied in the referential of the growing economy.

\section{Motivations and organization of the book}

\begin{figure}[h]
\includegraphics[width=0.5\textwidth]{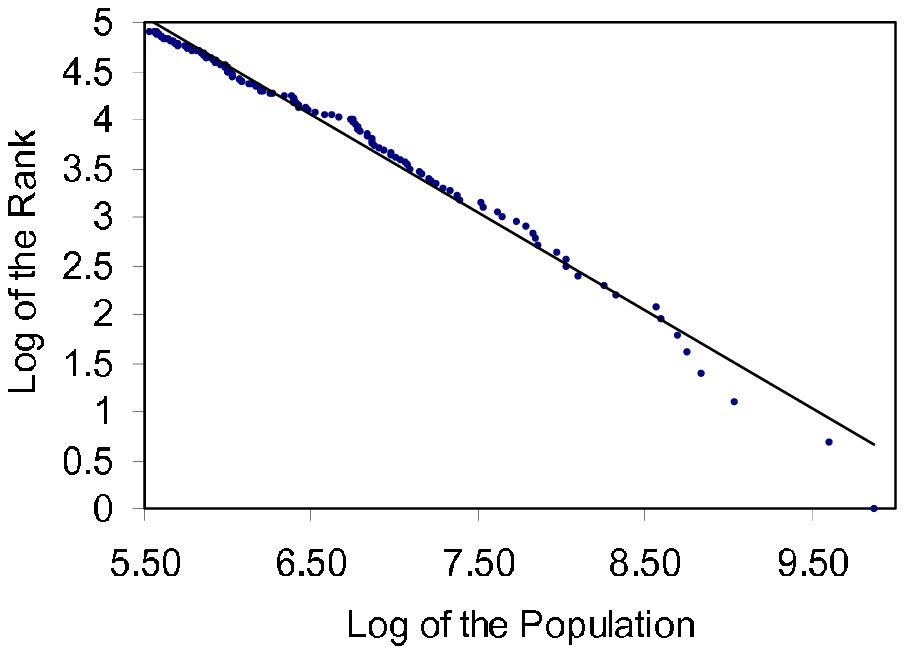}
\includegraphics[width=0.5\textwidth]{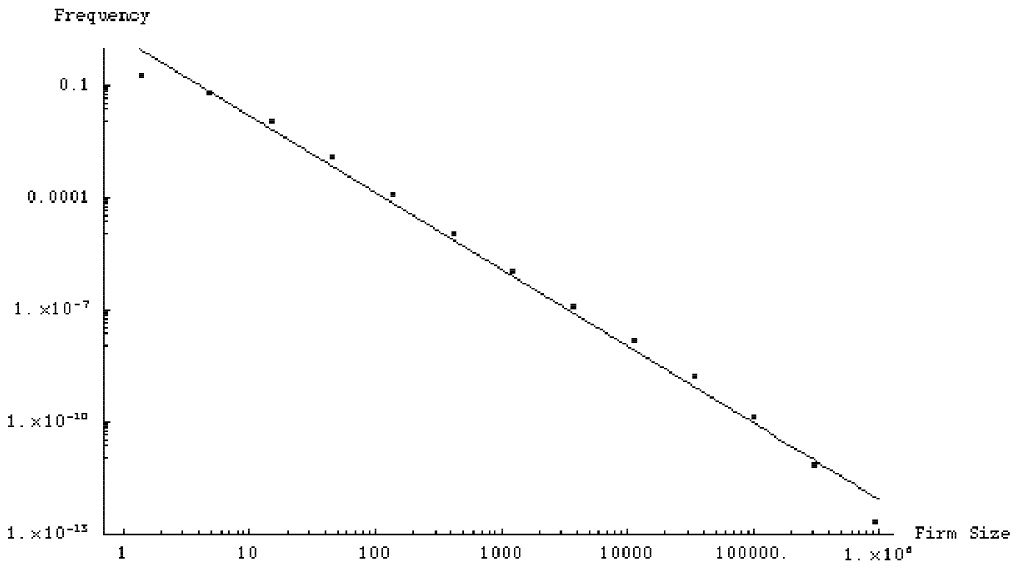}
\includegraphics[width=0.5\textwidth]{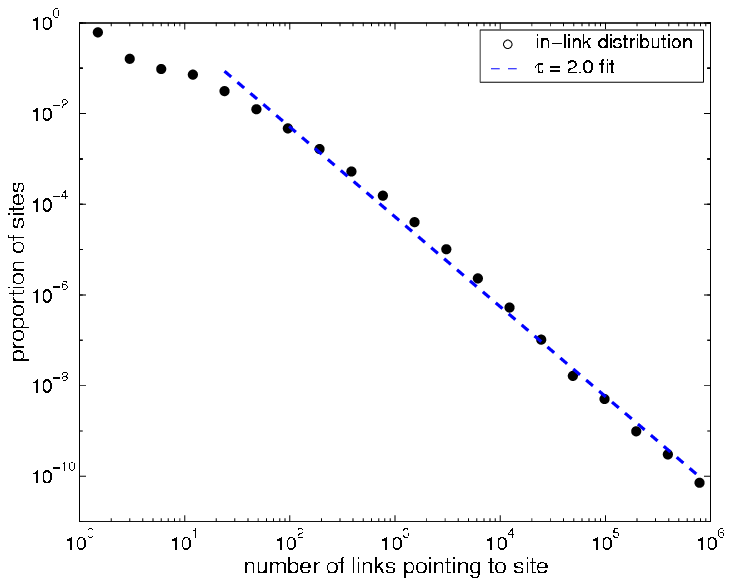}
\includegraphics[width=0.5\textwidth]{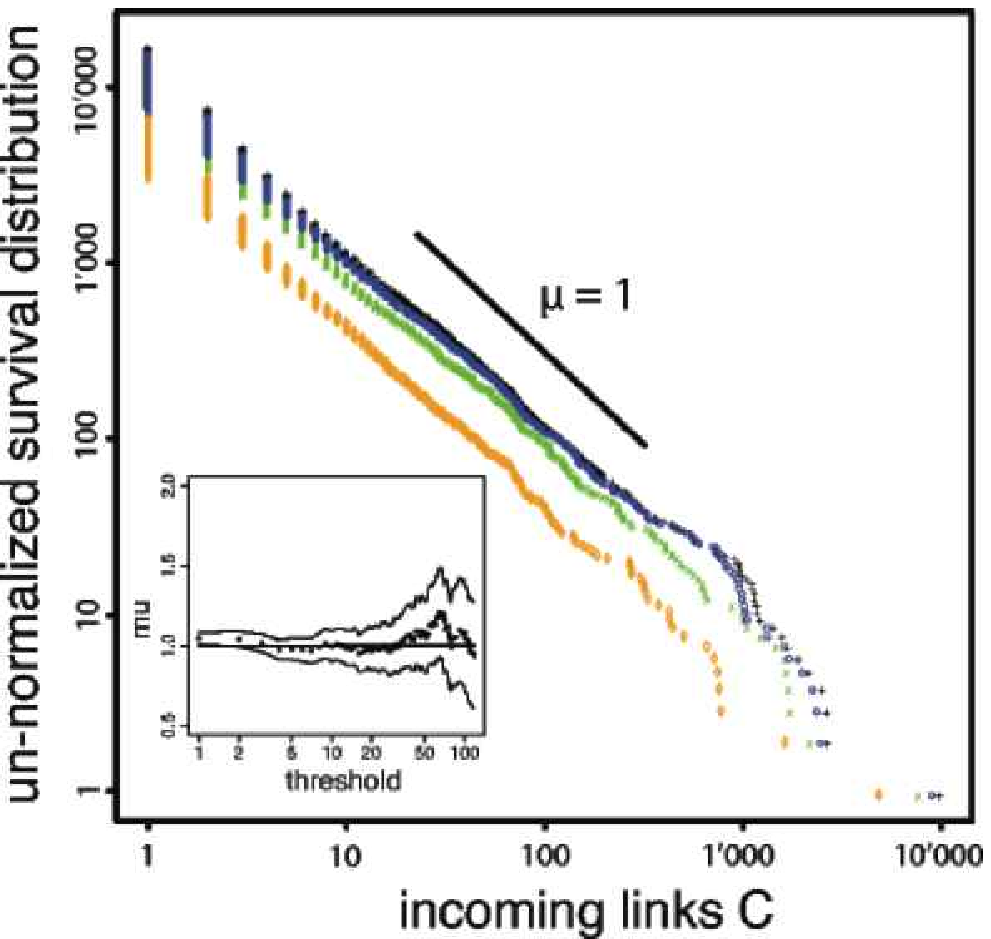}
\caption{ \label{FigExamples} Illustration of Zipf's law for city sizes (upper left panel, reproduced from \protect\citeasnoun{IoannidesGabaix03}), for firm sizes (upper right panel, reproduced from \protect\citeasnoun{Axtell01}), for the number of Internet links pointing to some website (lower left panel, reproduced from \protect\citeasnoun{Huberman2}) and for the number of incoming links to packages found in different Linux open source software releases (lower right panel, reproduced from \protect\citeasnoun{MSSVK08}).}
\end{figure}

One of the broadly accepted universal laws of complex systems, particularly
relevant in social sciences and economics, is that proposed by \citeasnoun{Zipf49}.
Zipf's law usually refers to the fact that the survival probability $P(s)=
\Pr\{S>s\}$ that the value $S$ of some stochastic variable, usually a size or
frequency,  is greater than $s$, decays with the growth of $s$ as $P(s)\sim
s^{-1}$. This in turn means that the probability density functions $p(s)$  exhibits
the power law dependence
\begin{equation}
p(s) \sim 1 / s^{1+m}~~~{\rm with}~~m = 1~. \label{wbtwr}
\end{equation}
Perhaps the distribution most studied from the perspective of Zipf's law is that of
firm sizes, where size is proxied by  sales, income, number of employees, or total
assets. Many studies have confirmed the validity of Zipf's law for firm sizes
existing at current time $t$ and estimated with these different measures
\cite{SimonBonini,IS77,Sutton97,Axtell01,Okuyama1999,Gaffeo03,Aoyama04,Fujiwara_etal_04a,Fujiwara_etal_04b}.
Initially formulated as a rank-frequency relationship quantifying the relative
commonness of words in natural languages \cite{Zipf49}, Zipf's law accounts
remarkably well for the distribution of city sizes \cite{Gabaix99} as well as firm
sizes all over the world, as just mentioned. Recently, Zipf's law has also been
found in Web access statistics and Internet traffic characteristics
\cite{Huberman1,AlbertBara02} as well as in bibliometrics, informetrics,
scientometrics, and library science (see \cite{Huberman2} and references therein).
There are also suggestions for applications to other physical and biological,
sociological and financial market processes (see list of references in
\url{http://linkage.rockefeller.edu/wli/zipf/index_ru.html}).  Figure~\ref{FigExamples} illustrates several applications of Zipf's law to different fields of social and natural sciences.

Among the many more or less successful explanations proposed to understand the
origin of Zipf's law, one of the  most promising is the explanation by
\citeasnoun{Gabaix99} and \citeasnoun{IoannidesGabaix03} formulated in the context
of the distribution of city sizes, based on Gibrat's law. \citeasnoun{Gabaix99}
assumed  that each city exhibits a stochastic growth rate distributed independently
from its present size. \citeasnoun{Gabaix99} showed that Gibrat's law for city
growth (together with some important deviations of Gibrat's law), normalized to the
whole population of a given country, leads to distributions of city sizes very
close to Zipf's law. However, the derivation of \citeasnoun{Gabaix99} suffers from
a few  problems.

First, the exact scale-independent Gibrat's law leads to a log-normal distribution
of city sizes, which is not strictly a power law and only slowly converges to a
power law in the limit of large log-variance (and some other conditions), becoming
at the same time more and more degenerate. Some additional assumptions are
therefore needed in order to produce the stable non-degenerate Zipf's law. In
particular, \citeasnoun{Gabaix99} assumed that, for cities of small sizes, there
are some exogenous factors preventing further decaying of their population (see
also \cite{LevySolomon96,Malcai_etal99}). More appropriate to social and economic
phenomena is the supposition, contrary to preventing population decay, of
eliminating cities or firms as they reach a small size. An example is the
transition from city to rank of village as the size goes below some threshold. In
the context of an economy of firms, it is important to take into account the
continuous process with births and deaths playing a central role at time scales as
short as a few years. A goal of the present book is to demonstrate that death (as
well as birth) processes are especially important to understand the economic
foundation of Zipf's law and its robustness. We will consider two different
mechanisms for the exit of a firm: (i) when the firm total asset value becomes
smaller than a given minimum threshold (which can vary with time and with
countries) and (ii) when an exogenous shock occurs, modeling for instance
operational risks, independently of the size of the firm.

Another shortcoming of Gabaix's approach is the simplifying supposition that all
cities originate at the same instant $t_0$, and then only grow stochastically,
obeying the balanced Gibrat's law mentioned above. We believe that it is more
realistic, especially for the description of the behavior of the asset value of
firms (which is more dynamic than the formation of cities), that the births of
firms occur according to a random point process characterized by some mean rate
$\nu(t)$. Jointly, one should take into account the well-documented evidence that
firms die, for instance when their size go under some low asset value level. It
turns out that taking into account the random flow of firm births and deaths, in
combination with Gibrat's law, leads to the pure and non-degenerate Zipf's law,
without the need to the rather artificial modification of Zipf's law for small
sizes [We note that the fact that deviation of Gibrat's law has been documented for
small firms is another issue, as the documented deviations do not necessarily obey
the assumptions needed in Gabaix's derivation.]  As a bonus, the approach in terms
of the dynamics of birth-death together with stochastic growth, that we develop
here, leads to specific predictions of the conditions under which deviations from
Zipf's law occur, which help rationalize the empirical evidence documented in the
literature. The conditions involve either deviations from Gibrat's law in the
stochastic growth process of firms or the existence of an unbalanced growth or
decay of the mean birth rate $\nu(t)$ of new firms, as we explain in details below.

For transparency of derivations and for convenience of analytic calculations, we
use a continuous version of Gibrat's law, allowing us to benefit from the
properties of the Wiener process and the mathematical framework of Kolmogorov's
diffusion equations. We unearth new properties associated with the stochastic
behavior of firm assets. We show that the death of firms at some low value level as
well as possibly significant deviations from Gibrat's law do not affect the
asymptotic validity of Zipf's law in the limit of large firm sizes. By analyzing a
large
 class of diffusion processes modeling the behavior of firm assets
with growth rates very different from Gibrat's condition,  we find general
conditions for the validity of Zipf's law. Specifically, we have discovered
stochastic growth models with non-Gibrat properties, leading to Zipf's and related
power laws for the current density of firms' asset values.

The  book is organized as follows. Chapter 2 presents the continuous version of
Gibrat's law and some peculiarities of the stochastic behavior of the geometric
Brownian motion of firms' asset values, resulting from Gibrat's law.

Chapter 3 describes the proposed model for the current density of firms' asset
values, taking into account the random flow of the birth of firms. We show that, if
some natural balance condition holds, which is analogous to \citeasnoun{Gabaix99}
normalizing condition, while the mean birth rate of firms is independent of time
($\nu=\text{const}$), then the exact Zipf's law holds true.

Amazingly, despite the relevance of Gibrat's law and the corresponding geometric
Brownian motion in a wide range of physical, biological, sociological and other
applications, many researchers do not make use of many of the interesting
properties exhibited by realizations of the geometric Brownian motion, in order to
derive detailed explanations of Zipf's and related power laws. Thus, in chapter~4,
we gather little-known information concerning the statistical properties of
realizations of the geometric Brownian motion, which play a significant role for
the understanding of the roots and conditions of the validity of Zipf's law.

Chapter~5   discusses in detail the influence  on the validity of Zipf's law of the
occurrence of the death of firms when their value falls below some low level. In
chapter~6, we derive an equation for the steady-state density of firm asset values,
which enables us to explore in detail the consequences of deviations from Gibrat's
law at moderate asset values on the validity of Zipf's law at higher asset values.

Chapters~7 and 8 are devoted to discussing possible deviations from Zipf's law due
respectively to the sudden death of firms and the time dependence of the birth
rate. It is shown that, even in such situations, Zipf's law holds if some
generalized balance condition is valid. In particular, we discuss the robustness of
Zipf's law to variations of the mean birth rate and of the rate of growth of the
mean asset value of particular firms. The second part of Section~8 presents a
simple coupled model describing the possible connection between the stochastic
behavior of firms' asset values and the mean birth rate.

In addition to the mechanisms in terms of birth, death and random growth which have been considered in the previous chapters, we envision that the next level of development of a complete mathematical theory of firms needs to take into account the mechanism of mergers between  firms (referred to as M\&A for ``merger and acquisition''), as well as it symmetric, the phenomenon of
creation of spin-off firms created from parent firms which privatize a part of their existing business as separate units.  For this, the long tradition in physics concerning the investigation of the processes of coagulation (merger) and of fragmentation (spin-off) could provide a fertile reservoir of ideas and techniques \cite{Aldous99,Leyvraz03}. Chapter 9 presents the integro-differential equation that expresses the coupling between firms introduced by M\&A and spinoffs and provides
preliminary results. This section is more an appetizer and encouragement for future works
than a complete treatment.

\section*{Table of contents}

2. Continuous Gibrat's law and Gabaix's derivation of Zipf's law

~~~2.1 Definition of continuous Gibrat's law

~~~2.2 Geometric Brownian motion

~~~2.3 Self-similar properties of the geometric Brownian motion

~~~2.4 Time reversible geometric Brownian motion

~~~2.5 Balance condition

~~~2.6 Log-normal distribution

~~~2.7 Gabaix's steady-state distribution 

3. Flow of firm creation

~~~3.1 Empirical evidence and previous works on the arrival of new firms 

~~~3.2 Mathematical formulation of the flow of firms' births at random instants

~~~3.3 Steady-state density of firms' asset values obeying Gibrat's law

~~~3.4 Quick and dirty explanation of the origin of the power law distribution of firm szies

4. Useful properties of realizations of the geometric Brownian motion

~~~4.1 Relationship between the distributin of firm's waiting times and sizes

~~~4.2 Mean growth versus stochastic decay

~~~4.3 Geometrically transparent definitions of stochastically
  decaying and growing processes

~~~4.4 Majorant curves of stochastically decaying geometric
  Brownian motion

~~~4.5 Maximal value of stochastically decaying geometric
  Brownian motion
  
~~~4.6 Extremal properties of realizations of stochastically
  growing geometric Brownian motion

~~~4.7 Quantile curves

~~~4.8 Geometric explanation of the steady-state density of a firm's asset value

5. Exit or ``death'' of firms

~~~5.1 Empirical evidence and previous works on the exit of firms

~~~5.2 Life-span above a given level

~~~5.3 Distribution of firms' life durations above a survival level

~~~5.4 Killing of firms upon first reaching a given asset level from above 

~~~5.5 Life-span of finitely living firms

~~~5.6 Influence of firms' death on the balance condition

~~~5.7 Firms' death does not destroy Zipf's law

6. Deviations from Gibrat's law and implications for generalized Zipf's laws

~~~6.1 Diffusion process with constant volatility

~~~6.2 Steady-state density of firms' asset values in the presence of deviations from Gibrat's law

~~~6.3 Integrated flow

~~~6.4 Semi-geometric Brownian motion

~~~6.5 Generalized semi-geometric Brownian motion

~~~~~~6.5.1 Statistic properties of generalized semi-GBM

~~~~~~6.5.2 Deterministic skeleton of the mean density of firm sizes

~~~~~~6.5.3 Stochastic equation for the process $S(t)=\omega[Y(t)]$

~~~~~~6.5.4 Size dependent drift and volatility

~~~6.6 Zipf's law for the steady-state density of firms'
asset values when the firm exit level is zero

~~~6.7 Zipf's laws when Gibrat's law does not hold

7. Firms' sudden deaths 

~~~7.1 Definition of the survival function 

~~~7.2 Exponential distribution of sudden deaths  

~~~7.3 Implications of the existence of sudden firm exits for semi-geometric Brownian motions

~~~7.4 Zipf's law in the presence of sudden deaths

~~~7.5 Explanation of the generalized balance condition

~~~7.6 Some consequences of  the generalized balance condition

~~~7.7 Zipf's law as a universal law with a large basin of attraction

~~~7.8 Rate of sudden death depending on firm's asset value

~~~7.9 Rate of sudden death depending on firm's age 

8. Non-stationary mean birth rate

~~~8.1 Exponential growth of firms' birth rate

~~~8.2 Deterministic skeleton of Zipf's law

~~~8.3 Mean density of firms younger than age $t$

~~~8.4 Simple model of birth rate coupled with the overall firms' value

~~~8.5 Dynamics of mean birth rate

9. Conclusions and future directions

~~~9.1 Importance of balance conditions for Zipf's law

~~~9.2 Mergers and acquisitions and spin-offs: general formalism

~~~9.3 Mergers and acquisitions and spin-offs with Brownian internal growth

~~~9.4 {Mergers and acquisitions and spin-offs with GBM for the internal growth process

\end{document}